# Kinetic Monte Carlo–Ising Machine Optimization for Atomistic Inverse Design of Solid Electrolytes

Ai Koizumi[1,2]*, Tomofumi Tada[3]*, Ryo Tamura[1,4]*

1 Center for Basic Research on Materials, National Institute for Materials Science, 1-1 Namiki, Tsukuba, Ibaraki 305-0044, Japan

2 Center for Computational Sciences, University of Tsukuba, 1-1-1 Tennodai, Tsukuba, Ibaraki 305-8577, Japan

3 Kyushu University Platform of Inter/Transdisciplinary Energy Research, Kyushu University, 744 Motooka, Nishi-ku, Fukuoka 819-0395, Japan

4 Graduate School of Frontier Sciences, The University of Tokyo, 5-1-5 Kashiwa-no-ha, Kashiwa, Chiba 277-8561, Japan

ABBREVIATIONS

Factorization machine with quadratic-optimization annealing; FMQA, quadratic unconstrained binary optimization; QUBO, kinetic Monte Carlo; KMC, 8 mol % yttria-stabilized zirconia; 8YSZ, solid oxide fuel cells; SOFCs

**ABSTRACT (150 words limit):**

Maximizing ionic conductivity remains a fundamental challenge in the atomistic design of solid electrolytes. To this end, we present a framework that combines kinetic Monte Carlo (KMC) and factorization machine with quadratic-optimization annealing (FMQA), an Ising-machine-based black-box optimization algorithm for large-scale combinatorial optimization. KMC evaluates the ionic conductivity for a given dopant configuration, whereas FMQA iteratively learns a surrogate model from a small configuration-conductivity dataset and proposes configurations expected to maximize conductivity. To address the severe combinatorial explosion in large KMC simulation cells, we partition the configuration space for parallel optimization. As a proof of concept, we apply this KMC–FMQA framework to bulk 8 mol % yttria-stabilized zirconia, identifying a dopant configuration with an order-of-magnitude higher conductivity than that of random configurations and the experimentally reported conductivity. Combined with experiments, this framework will enable the determination of microscopic structures from measured conductivity, providing insight into the underlying transport mechanisms.

**KEYWORDS:** Factorization machine with quadratic-optimization annealing (FMQA), Ising machine, kinetic Monte Carlo (KMC), ionic conductivity, solid oxide fuel cell (SOFC)

Solid oxide fuel cells (SOFCs)[1–6] are next-generation energy conversion devices, and ionic conductivity of solid electrolytes, which are a key component of SOFCs, has been extensively investigated to improve their performance. In the design of solid electrolytes, the microscopic structure plays a crucial role in determining ionic conductivity, especially when SOFC devices are miniaturized. For example, yttria-stabilized zirconia (YSZ) is a representative solid electrolyte used in SOFCs, and its ionic conductivity has been well investigated as a function of dopant concentration, and 8 mol % YSZ (8YSZ) exhibits the highest ionic conductivity.[7] In bulk 8YSZ, the dopants are randomly distributed, and the effect of the configuration on ionic conductivity is generally negligible. In contrast, thin film 8YSZ exhibits a wide range of ionic conductivity[8], suggesting a strong dependence on various microscopic structural features, including dopant configuration. Thus, if microscopic structures can be well controlled, the ionic conductivity of 8YSZ could be further enhanced. However, elucidating this microscopic structure–conductivity relationship experimentally remains challenging, motivating computational approaches to investigate it. Recently, we used kinetic Monte Carlo (KMC) simulations[9–11] to investigate drift currents arising from anisotropic oxygen-ion diffusion under an applied bias voltage while varying the dopant configuration in bulk 8YSZ, thereby demonstrating that dopant configurations have a profound effect on ionic conductivity. Furthermore, we established a link between microscopic dopant configurations and ionic conductivities using a structural descriptor related to ion-migration paths.[11] In other words, ionic conductivity was explicitly shown to depend systematically on the microscopic dopant configuration.

**Figure 1**. KMC–FMQA framework. In (a), the factorization machine (FM) is trained using the relationship between the dopant configurations encoded as bit strings and their ionic conductivities, and the trained FM is used by an Ising machine (b) to explore promising dopant configurations. KMC simulations are performed on the systems corresponding to the proposed bit configurations in (c), and the resulting ionic conductivities are added to the training dataset for the next cycle. Green and gray spheres represent zirconium (Zr) and yttrium (Y) lattice ions to be optimized,

respectively, whereas red spheres represent oxygen ions migrating through paths consisting of six lattice sites via oxygen vacancies (d).

The ultimate goal is not to link dopant configurations to ionic conductivity, but to identify the configuration that maximizes ionic conductivity. This is the essence of atomistic inverse design.[12–14] However, since KMC only solves the forward problem of predicting ionic conductivity from a given configuration, achieving inverse design requires efficient exploration of the vast configuration space. To this end, black-box optimization (BBO)[15] is particularly well studied. BBO uses a surrogate model constructed from a small configuration-conductivity dataset to iteratively propose and evaluate configurations, thereby identifying configurations with the desired ionic conductivity without requiring knowledge of the full configuration–conductivity relationship. That said, identifying optimal atomic configurations in large multicomponent crystalline systems is challenging due to the combinatorial explosion in the number of possible configurations, which renders conventional BBO methods such as Bayesian optimization[16,17] intractable. In such situations, Factorization Machine with Quadratic-optimization Annealing (FMQA)[18,19] provides a powerful alternative, combining the efficiency of BBO with the combinatorial search capability of Ising machines[20]. Recently, we applied FMQA with Vienna Ab initio Simulation Package (VASP) to determine stable atomic configurations in a small supercell[21] and in a relatively large grain boundary system[22], thereby elucidating their atomic configurations. FMQA has thus emerged as a key technology in BBO, with successful applications ranging from crystal structure prediction to broader combinatorial configuration problems.[23–26]

Here, we demonstrate that the atomistic inverse design problem for solid electrolytes can be solved by combining KMC[9–11] with FMQA,[18,19] establishing an efficient structural design framework (KMC–FMQA framework), as illustrated in Fig. 1. KMC evaluates the ionic current for a given dopant configuration, and FMQA iteratively learns a surrogate model from a small configuration-current dataset and proposes configurations predicted to exhibit the desired ionic current. We apply this framework to optimize the dopant configuration in a 16-layer region of a bulk 8YSZ supercell. For efficient exploration of the vast configuration space, we introduce a block model in which four lattice sites are grouped into a single unit (block), and three types of blocks, $Zr_4$, $Y_2Zr_2$, and $Y_4$, are used to construct the 8YSZ supercell (see Fig. 1(a)). To maintain the 8YSZ dopant concentration, we enumerated all 81 possible block compositions ($b$=0-80), as shown in Figure 2(a). We then partitioned the search space into 81 compositional subspaces and explored each subspace independently. In this letter, we set the inverse-design target to identify the block configuration exhibiting the highest ionic conductivity. Further details are described in Methods.

Figure 2(b) shows the evolution of the highest ionic conductivity obtained in 81 parallel KMC–FMQA optimization runs over 30 steps with 10 initial training data. FMQA consistently identified dopant configurations with higher ionic conductivity than those obtained by the initial random sampling, demonstrating the effectiveness of the KMC-FMQA framework. Note that histograms of the ionic conductivities are shown in Fig. S1. To further explore promising configurations, an additional 70 optimization steps were performed for the three block compositions with $b$ = 20, 14, and 8, that exhibited the highest ionic conductivities during the first 30 steps (see Fig. S2). The optimization trajectories varied significantly depending on the initial dataset, as shown in Fig. S2, suggesting that a strategy for initial data selection is necessary to ensure more robust optimization.

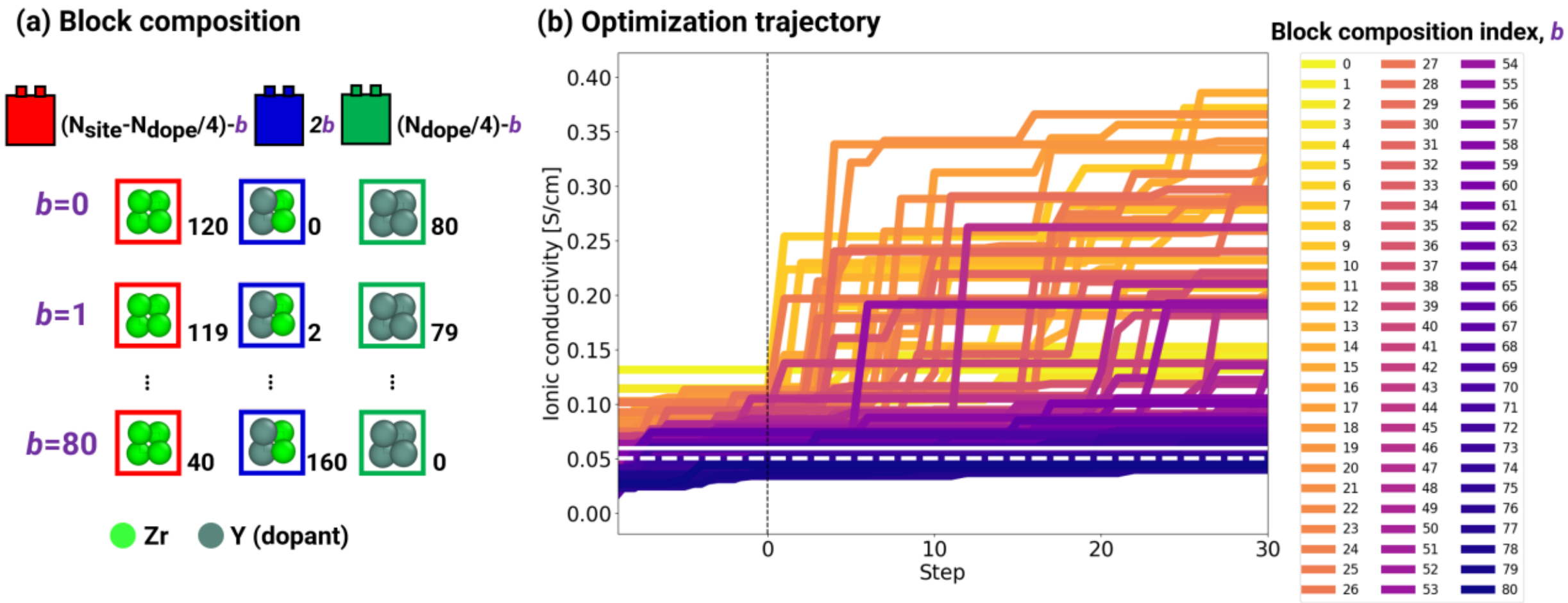

**Figure 2**. (a) Eighty-one block compositions satisfying the 8YSZ dopant concentration, where $N_{\mathrm{site}}$ and $N_{\mathrm{dope}}$ denote the numbers of block placement sites and yttrium (Y) dopants in the system, respectively ($N_{\mathrm{site}}$ = 200, $N_{\mathrm{dope}}$ = 320). $b$ denotes the block composition index. (b) Optimization trajectories for each of the 81 block compositions. Each trajectory shows the highest ionic conductivity achieved at each iteration step. The ionic conductivity up to step 0 corresponds to the initial training data obtained from 10 evaluations through random sampling, and the subsequent values are obtained through FMQA optimization. The white solid and white dashed horizontal lines represent the average conductivity for random dopant distributions[11] and the experimentally reported conductivity of bulk 8YSZ[27], respectively.

The optimization identified a dopant configuration with an ionic conductivity of 0.423 S cm$^{-1}$, which is an order-of-magnitude higher than both the average conductivity for random dopant distributions (0.06 S cm$^{-1}$)[11] and the experimentally reported conductivity of bulk 8YSZ (0.0505 S cm$^{-1}$).[27] Figures 3(a)-(c) show the three highest-conductivity configurations for the three block compositions with $b$ = 20, 14, and 8, whereas Figs. 3(d)-(f) show the configurations with ionic conductivity close to the experimentally reported value, selected from the initial random sampling datasets for the same block compositions. In the high-conductivity configurations, dopants are highly concentrated near the edges of the 16 layers (i.e., at the YSZ/$ZrO_2$ interfaces). The ease of ion migration is determined by the dopant configuration in the six-site ion-migration path shown in Fig. 1(d) and by the electric field (see our previous study[11] for details). When dopants are locally concentrated in the path, the configuration-dependent migration barrier tends to be higher (see Table S2 in our previous study[11]), whereas layered dopant configurations shown in Figs. 3(a)-(c) promote oxygen vacancy accumulation along the layers, which modifies the electrostatic contributions to the migration barriers and may thereby facilitate ionic transport. To fully elucidate the configuration–conductivity relationship, further investigations are required, including exploring a broader range of dopant configurations and identifying key structural features from high-dimensional structural descriptors governing ionic conductivity[11].

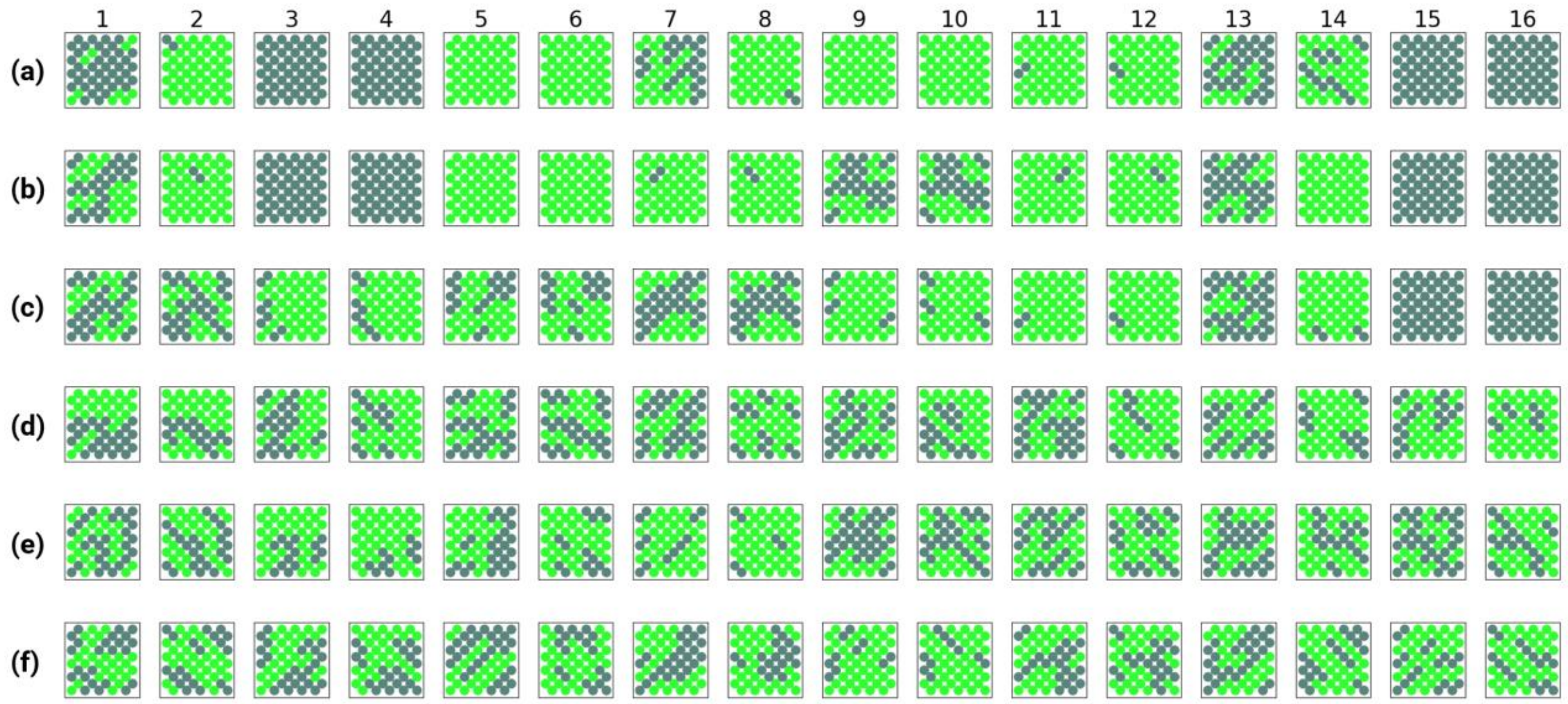

**Figure 3**. Cross-sectional images of high-conductivity configurations sliced into 16 layers. (a)-(c) The three highest-conductivity configurations identified by the KMC-FMQA framework, with ionic conductivities of 0.423, 0.400, and 0.371 S $cm^{-1}$, respectively. (d)-(f) Configurations with ionic conductivity close to the experimentally reported value, selected from the initial random

datasets corresponding to (a)-(c). Green and gray spheres represent zirconium (Zr) and yttrium (Y) atoms, respectively.

To benchmark the framework, we compared three approaches with progressively reduced search spaces: the all-atom model, the block model without the composition-based partitioning based on the block composition index $b$, and the block model with the partitioning (see Supporting Note A). The block model with composition-based partitioning achieved the best optimization performance among the three approaches, demonstrating that a block-based representation combined with search-space partitioning is more effective than simply increasing the number of trials in a vast search space.

In summary, as a proof of concept, we demonstrated that combining KMC with FMQA enables the atomistic inverse design of solid electrolytes, identifying a dopant configuration with an order-of-magnitude higher ionic conductivity than the experimentally reported conductivity of bulk 8YSZ. The framework efficiently explores the vast configuration space through a block-based representation combined with composition-based partitioning, enabling rapid identification of high-conductivity configurations. In the future, by minimizing the discrepancy between experimental and calculated conductivities, the framework will enable inverse estimation of microscopic structures, potentially providing deeper microscopic insight into ionic transport mechanisms in SOFC electrolytes. The approach is readily extensible to electrolyte-electrode interfaces, opening a pathway toward broader atomistic materials design for SOFC performance optimization.

# METHODS

An overview of the KMC–FMQA framework is summarized in Fig. 1. KMC evaluates the ionic conductivity for a given dopant configuration, whereas FMQA learns a surrogate model from a small configuration-conductivity dataset and optimizes the dopant configuration to maximize ionic conductivity. Here, we first briefly explain the KMC evaluation and then describe the FMQA optimization.

In KMC part, for given dopant configurations, KMC evaluates drift currents arising from mobile light oxygen-ions migrating (Fig. 1(d)), where the migration barriers mainly depend on the dopant configuration of the crystalline system. In contrast, FMQA updates dopant configurations to achieve the desired current. That is, KMC is employed to simulate the time evolution of light mobile-ions, whereas FMQA updates the configuration of heavy lattice-atoms. When the optimization targets energetic stability rather than ionic conductivity, the roles of KMC and FMQA become analogous to those of MD and MC in MC–MD algorithms[28], respectively. The KMC computational conditions are summarized in Table S1 and are described in detail in our previous study[11]. In the previous study, we observed diverse ionic conductivities when dopants were highly concentrated in 16 layers along an applied bias voltage direction shown in Fig. 1(c). Therefore, in this study, the KMC–FMQA framework was applied to the optimization of dopant configurations in the 16 layers between the dashed lines in Fig. 1(c).

To efficiently handle the large configuration space, we introduced a block model based on the six-site ion-migration paths that govern ionic conductivity[11] (see Supporting Note B for details). This representation significantly reduced the search space, yielding a Quadratic Unconstrained Binary Optimization (QUBO) model that can be efficiently solved using an Ising machine. The 16 layers were constructed using three types of blocks (see Fig. 1(a)), and 81 block compositions satisfying the 8YSZ dopant concentration were enumerated in advance (see Fig. 2(a)). The search space reduction is provided in Supporting Note C.

FMQA was employed to optimize the block configuration to maximize ionic conductivity. Since only binary variables (0/1) can be used as inputs for FMQA, the block configurations were represented as bit structures; the three blocks were represented as 100, 010, and 001, as illustrated in Fig. 1(a). The bit $q_{it}$ satisfies the following two constraints:

$$\sum_{t=1}^{3} q_{it} = 1, \forall i, \tag{1}$$

where $i$ and $t$ denote the site and block-type indices, respectively, and when constructing the system by a block composition with index $b$,

$$\sum_{i=1}^{N_{\mathrm{site}}} q_{it} = N_t(b), \tag{2}$$

where $N_t(b)$ denotes the number of blocks of type $t$, and $N_{\mathrm{site}}$ denotes the number of block placement sites. The first constraint ensures that each site is assigned exactly one block type (one-hot constraint), and the second fixes the block composition to satisfy the 8YSZ dopant concentration. The optimization was performed as following four steps.

1. Initial Training Data Generation: 10 block configurations were generated by random sampling under the two constraints of Eqs. (1) and (2), and KMC simulations were performed to obtain the corresponding ionic conductivities.

2. Model Training: The factorization machine (FM) was trained using the initial dataset. The ionic conductivity prediction model is defined as follows.

$$\sigma(\mathbf{q}) = \sum_{i=1}^{N_{\mathrm{site}}} \sum_{t=1}^{3} w_{it}^{*} q_{it} + \sum_{i=1}^{N_{\mathrm{site}}} \sum_{t=1}^{3} \sum_{j=1}^{N_{\mathrm{site}}} \sum_{u=1}^{3} \sum_{k=1}^{K} v_{itk}^{*} v_{juk}^{*} q_{it} q_{ju}\,, \tag{3}$$

where $\mathbf{q}$ denotes a block configuration represented as a bit structure, $w_{it}^{*}$ and $v_{itk}^{*}$ denote the trained parameters, and $K$ denotes the dimension of the latent space (hyperparameter) which was set to 8 (default value)[29].

3. Ising Machine Optimization: A promising configuration $\mathbf{q}$ that maximizes the trained FM is determined by an Ising machine under the two constraints of Eqs. (1) and (2). Fixstars Amplify AE[30] was employed as the Ising machine. The most promising solution, $\mathbf{q}^{*}$, was selected in each cycle. If $\mathbf{q}^{*}$ violated the constraints or was already included in the training data, $\mathbf{q}^{*}$ was regenerated by random sampling.
4. Ionic Conductivity Calculation: The KMC simulation is performed on the system identified by $\mathbf{q}^{*}$. The obtained conductivity is added to the training data, and the procedure then returns to Step 2.

The crystalline system exhibits high symmetry, and 100 symmetry operations were applied (see Supporting Note D), thereby expanding the training dataset by a factor of 100 from a single KMC run. In this study, for each of the 81 block compositions, we performed three independent optimization trials with different initial training datasets. We then determined the highest conductivity in three independent trials and plotted it as a function of optimization step in Fig. 2(b).

**ACKNOWLEDGMENT**

This study was supported by the Core Research for Evolutional Science and Technology (JST CREST) (JPMJCR2234). T.T. acknowledges the support of JSPS KAKENHI (24H02203). The authors would like to thank Prof. Naoki Ohashi, Prof. Takashi Washio and Prof. Tomoyuki Higuchi for valuable discussions and insightful comments.

ASSOCIATED CONTENT

**Supporting Information.**

KMC condition; Comparison of ionic conductivity histograms from random sampling and FMQA; Optimization trajectories with three highest ionic conductivities; Optimization performance under different constraints; Block model; Search space reduction; Symmetry operations

AUTHOR INFORMATION

**Corresponding Authors**

Ai Koizumi – Center for Basic Research on Materials, National Institute for Materials Science (NIMS), 1-1 Namiki, Tsukuba, Ibaraki 305-0044, Japan; Center for Computational Sciences, University of Tsukuba, 1-1-1 Tennodai, Tsukuba, Ibaraki 305-8577, Japan; ORCID: https://orcid.org/0000-0003-4609-4586; Email: koizumi.ai@ccs.tsukuba.ac.jp

Tomofumi Tada – Kyushu University Platform of Inter/Transdisciplinary Energy Research, Kyushu University, 744 Motooka, Nishi-ku, Fukuoka 819-0395, Japan; ORCID: https://orcid.org/0000-0003-3093-3779; Email: tada.tomofumi.054@m.kyushu-u.ac.jp

Ryo Tamura – Center for Basic Research on Materials, National Institute for Materials Science (NIMS), 1-1 Namiki, Tsukuba, Ibaraki 305-0044, Japan; Graduate School of Frontier Sciences, The University of Tokyo, 5-1-5 Kashiwa-no-ha, Kashiwa, Chiba 277-8561, Japan; ORCID: https://orcid.org/0000-0002-0349-358X; Email: TAMURA.Ryo@nims.go.jp

TOC:

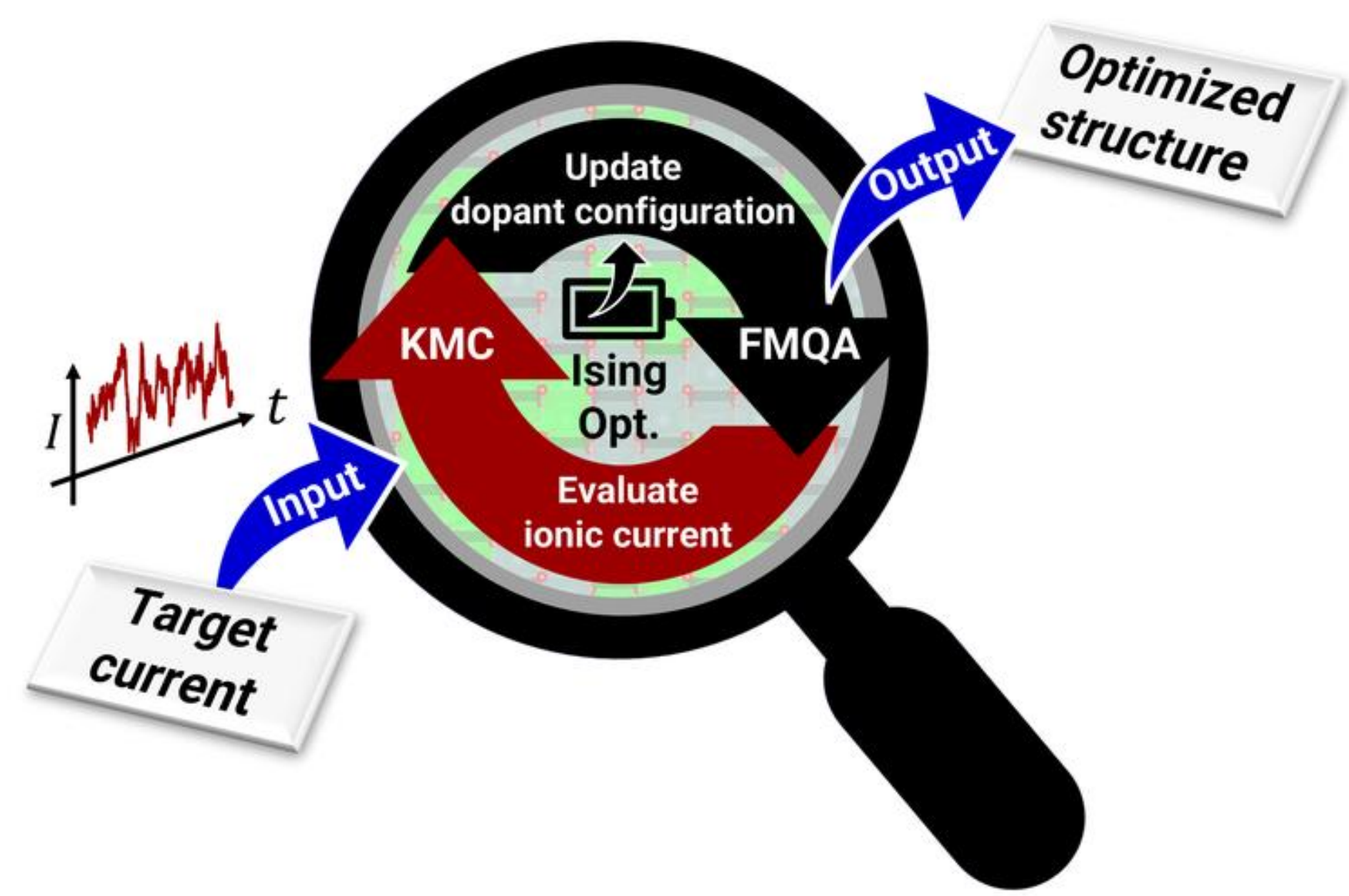

Update
dopant configuration
KMC
Ising
Opt.
FMQA
Evaluate
ionic current
I
t
Input
Output
Target
current
Optimized
structure

# Supporting Information for "Kinetic Monte Carlo–Ising Machine Optimization for Atomistic Inverse Design of Solid Electrolytes"

Ai Koizumi[1,2]*, Tomofumi Tada[3]*, Ryo Tamura[1,4]*

1 Center for Basic Research on Materials, National Institute for Materials Science, 1-1 Namiki, Tsukuba, Ibaraki 305-0044, Japan

2 Center for Computational Sciences, University of Tsukuba, 1-1-1 Tennodai, Tsukuba, Ibaraki 305-8577, Japan

3 Kyushu University Platform of Inter/Transdisciplinary Energy Research, Kyushu University, 744 Motooka, Nishi-ku, Fukuoka 819-0395, Japan

4 Graduate School of Frontier Sciences, The University of Tokyo, 5-1-5 Kashiwa-no-ha, Kashiwa, Chiba 277-8561, Japan

**Email:** koizumi.ai@ccs.tsukuba.ac.jp, tada.tomofumi.054@m.kyushu-u.ac.jp, TAMURA.Ryo@nims.go.jp.

**Table S1** *KMC Condition for calculating ionic conductivity.* See Table S3 in our previous study[1] for other detailed simulation conditions.

| | |
|---|---|
| Temperature | 1050 K |
| Number of equilibration steps | 1000 |
| Number of sampling steps | 8000000 |
| Dielectric constant | 27 |
| Bias voltage | 0.0002 V/Å |

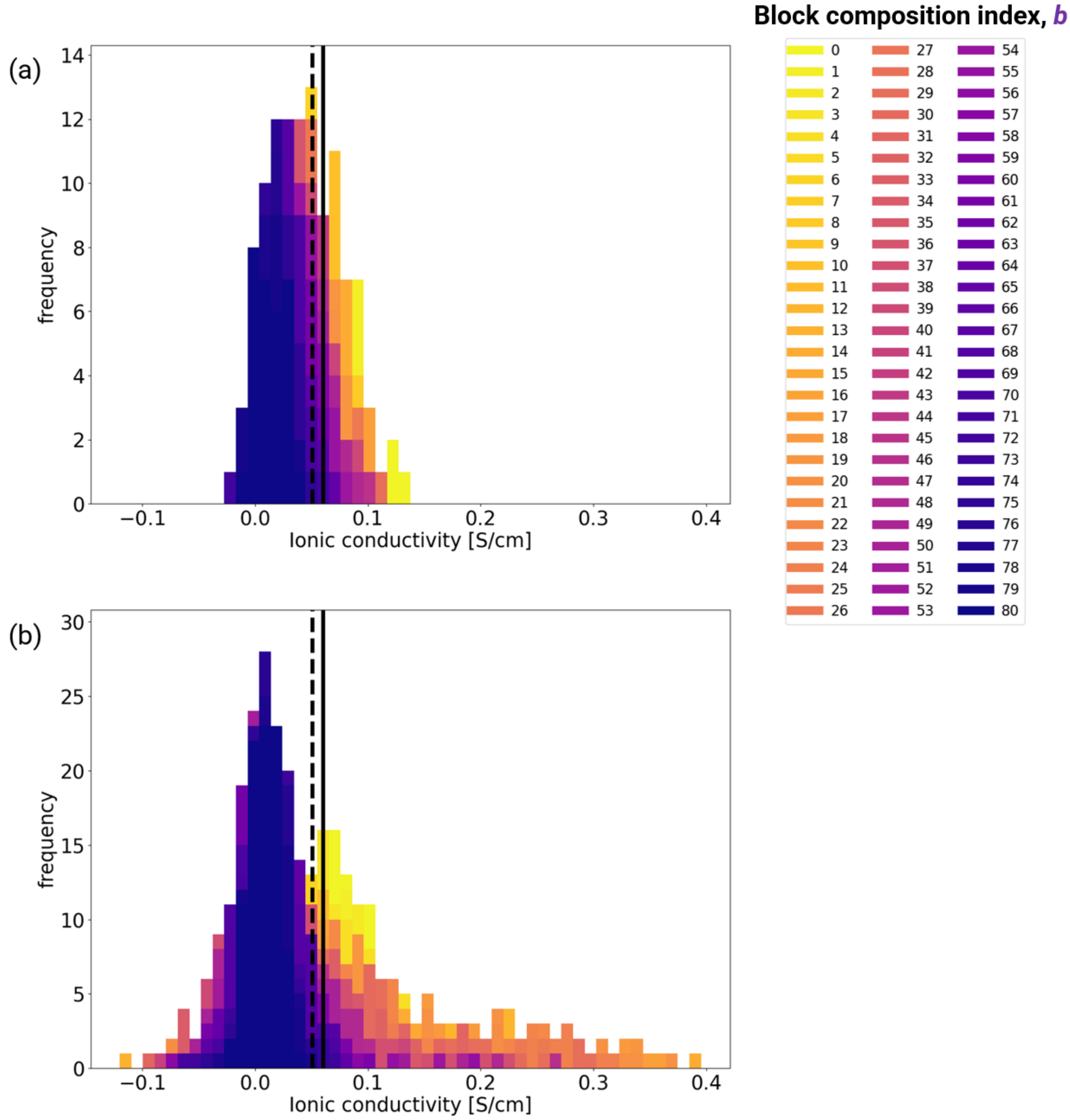

**Figure S1** *Comparison of ionic conductivity histograms from random sampling (a) and FMQA (b).* Both panels (a) and (b) show 81 histograms, one for each of the 81 block compositions; the composition index $b$ is defined in Fig. S4(e). The 81 histograms in (a) were obtained using the initial training data generated through random sampling ($N$=10×3 for each block composition), whereas those in (b) were obtained using data generated through Ising optimization ($N$=30×3 for each block composition). The black solid and black dashed vertical lines represent the average conductivity ($N$=100) for random dopant distributions across the entire system (0.06 S $cm^{-1}$)[1] and the experimentally reported conductivity of bulk 8YSZ (0.0505 S $cm^{-1}$)[2], respectively.

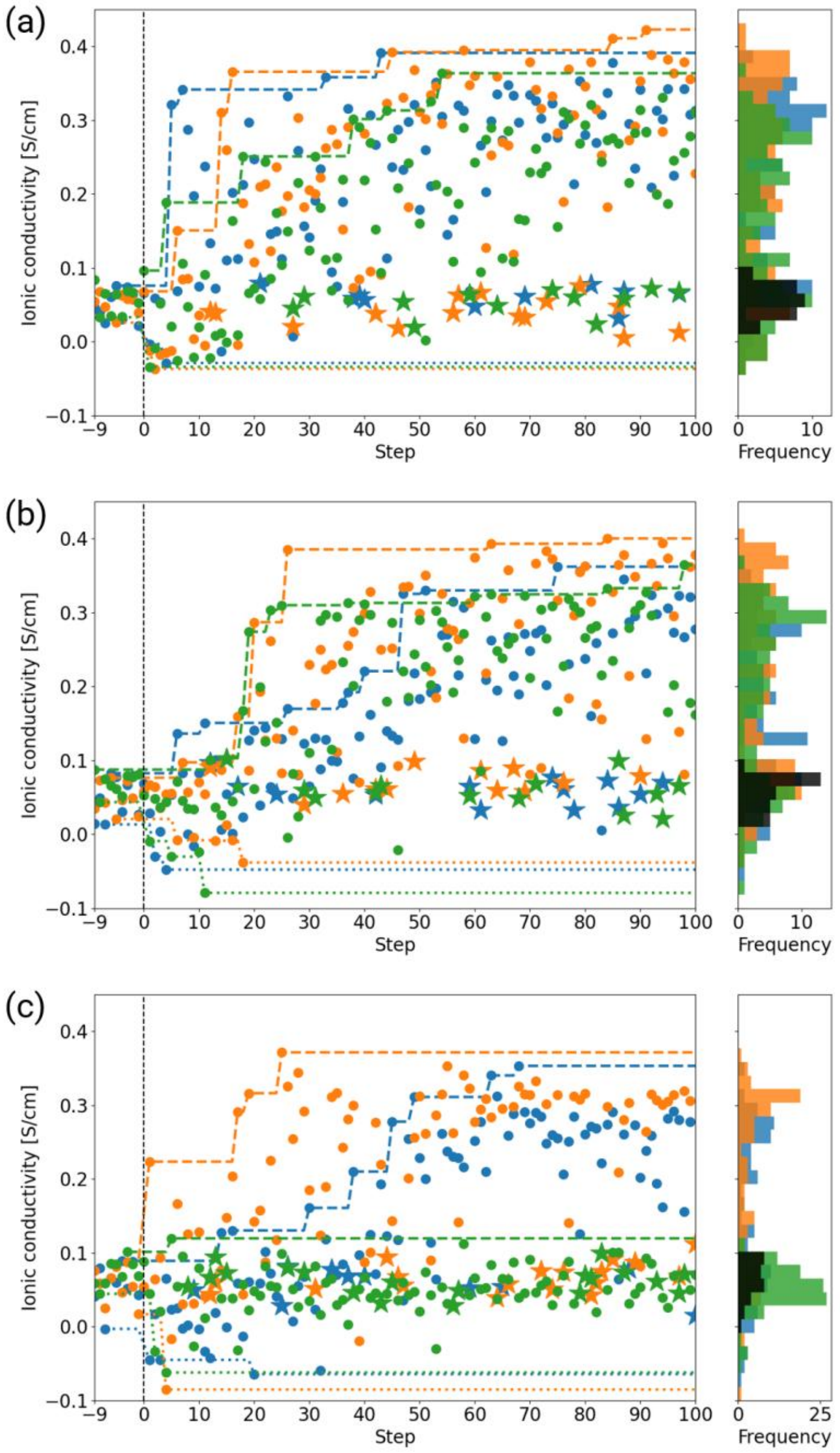

**Figure S2** *Three independent optimization trajectories for each of the three block compositions exhibiting the highest conductivities.* Three independent trials are shown in different colors. These three block compositions correspond to $b$ = 20 (a), 14 (b), and 8 (c). Points indicate the ionic conductivities obtained at each step. Star markers represent ionic conductivities of the configurations generated through random sampling when FMQA failed to generate a novel or feasible candidate. The histograms of ionic conductivities for each trial are also plotted on the right side. The black histogram was created using all the initial training data up to the 0th step of the three independent trials.

**Supporting Note A: Optimization performance under different constraints**

In this study, we evaluated the optimization performance of the block model with block-composition-based partitioning (BWP), discussed in the main text, as well as that of the block model without such partitioning (BWOP) and the all-atom model (AA). In the case of AA, the dopant configuration can be represented using 0 (Zr) and 1 (Y), and the optimization requires only the concentration constraint:

$$\sum_{i}^{N_{\mathrm{site}}} q_i = N_{\mathrm{dope}}. \tag{1}$$

In contrast, for BWOP, the following two constraints, the one-hot constraint and the concentration constraint, are required, as in BWP.

$$\sum_{t}^{3} q_{it} = 1, \forall i, \tag{2}$$

$$\sum_{i}^{N_{\mathrm{site}}} \sum_{t}^{3} N_t^{\mathrm{Y}} q_{it} = N_{\mathrm{dope}}, \tag{3}$$

where $N_t^{\mathrm{Y}}$ denotes the number of Y atoms constructing the block of type $t$. The upper panel of Figure S3 shows the optimization trajectories for the two cases, compared with the 81 parallel explorations (i.e., BWP) in the background in gray, whereas the panels (AA) and (BWOP) of Fig. S3 show the trajectories of three independent trials. Although BWOP has a smaller search space than AA, the Ising optimization did not work for BWOP in any of the three trials. Generally, imposing many constraints makes optimization difficult. In both BWP and BWOP, an additional one-hot constraint is required, which may have contributed to the failure of Ising optimization in BWOP. Nevertheless, the 81 parallel explorations (BWP) exhibit the best performance, thereby enabling the rapid identification of microscopic structures with the desired current.

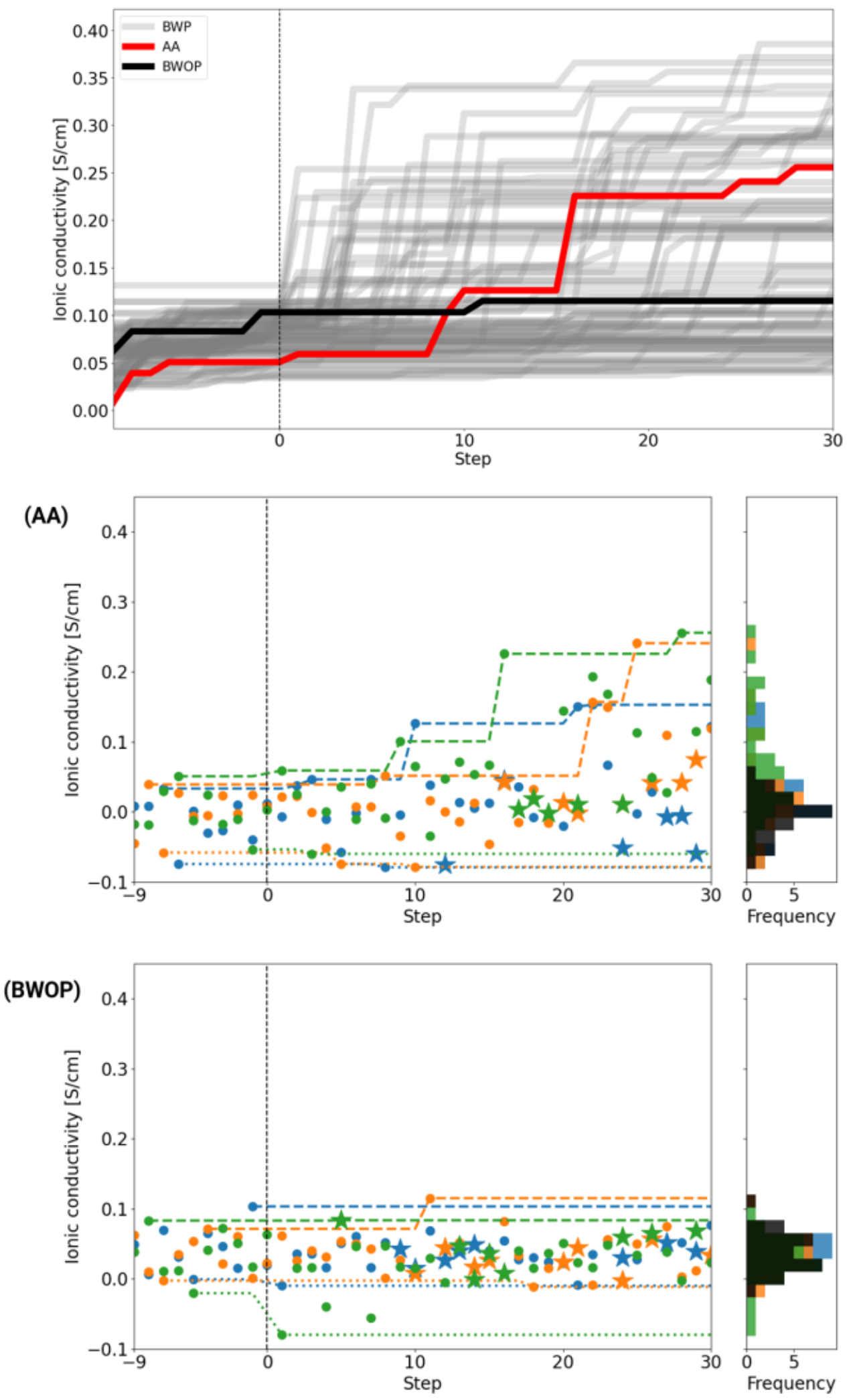

**Figure S3** *Optimization trajectories of the two additionally investigated cases.* (Upper panel) Trajectories show the highest ionic conductivity achieved at each iteration step. The highest conductivity of each step was determined in three independent trials with different initial training datasets. The ionic conductivity up to the 0th step corresponds to the initial training data generated through random sampling, whereas the subsequent values are obtained through the Ising optimization. The results of the 81 parallel explorations are also shown in the background in gray for comparison. (Bottom two panels) Optimization trajectories for AA and BWOP. Three independent trials are shown in different colors. Points represent the ionic conductivities obtained at each step. Star markers represent the ionic conductivities of the configurations generated through random sampling when the FMQA failed to generate a novel or a feasible candidate. The histograms of ionic conductivities for each trial are also plotted on the right side. The black histogram was created using all the initial training data from the three independent trials.

**Supporting Note B: Block model**

We explain the motivation for introducing the block model and the considerations that guided its modeling. In our previous study[1], based on 64 types of six-site ion-migration paths (see Fig. S4(a)), we defined the histograms of the paths that construct the system, and we linked the dopant configurations to ionic conductivity using the histograms. That is, the dopant configurations on the six sites (A–F) along the ion-migration path are important to characterize ionic conductivity. We thus believe that effective control of the ionic conductivity can be achieved by varying the composition and configuration of the 64 paths in the system. To reduce the search space, eight dopant configurations with representative migration energies were selected (see Fig. S4(b)); these configurations exhibit a wide range of migrtion energies (see the red star markers in Fig. S4(c)). As shown in Fig. S4(d), these eight configurations were decomposed into three types of four-site blocks, $Zr_4$, $Y_2Zr_2$, and $Y_4$. In this study, we used the blocks to construct the 16 layers. To fix the dopant concentration at 8YSZ, we enumerated the 81 block combinations in advance, as shown in Fig. S4(e). For example, the first block composition with $b = 0$ consists of 120 $Zr_4$ blocks, 0 $Y_2Zr_2$ blocks, and 80 $Y_4$ blocks, and the second with $b = 1$ consists of 119 $Zr_4$ blocks, 2 $Y_2Zr_2$ blocks, and 79 $Y_4$ blocks, that is, $b$ corresponds to half the number of $Y_2Zr_2$ blocks in each composition. The optimization was performed for each of 81 block compositions.

Here we briefly discuss the effect of block compositions on ionic conductivity. Figure S5 shows the ionic conductivity as a function of $b$; each data point represents the average ionic conductivity obtained from initial random sampling (N=10×3) for each block composition. Ionic conductivity decreases as $b$ increases, indicating that the $Y_2Zr_2$ blocks reduce ionic conductivity. Despite all simulation models being constructed with the same dopant concentration, we found that the ionic conductivity strongly depends on the block composition.

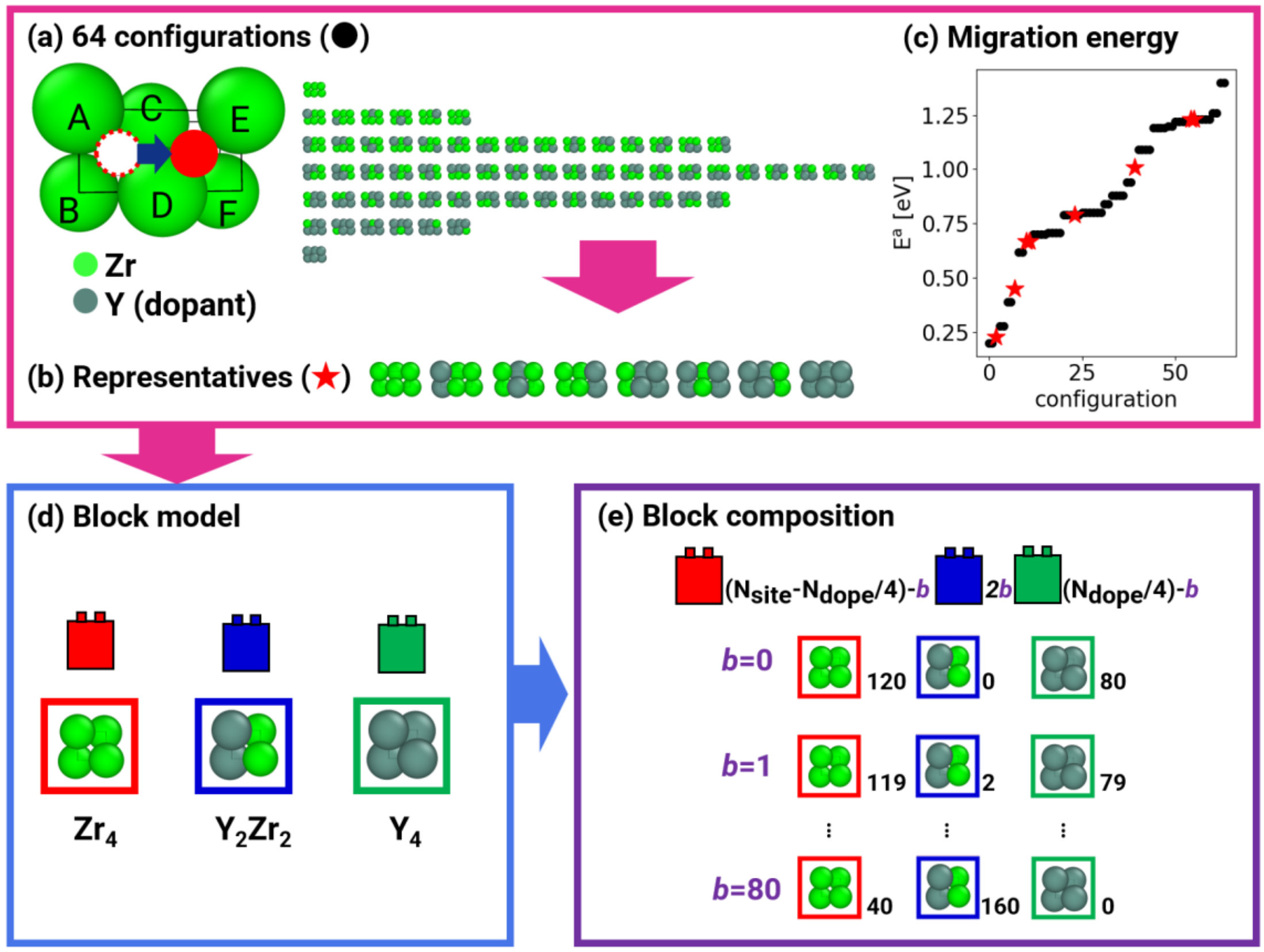

**Figure S4** *Schematic illustration of the thought process leading to block model.* (a) All 64 possible dopant configurations on the six sites (A–F) along the ion-migration path in the all-atom model. (b) Eight selected representative configurations. (c) Migration energies of all 64 configurations. The red star markers show the migration energies of the representatives. (d) *Block model*: The representatives were decomposed into three types of four-site blocks. (e) *Block composition*: Eighty-one block combinations satisfying the 8YSZ dopant concentration were used to construct the 16 layers. Green and gray spheres represent zirconium (Zr) and yttrium (Y) atoms, respectively.

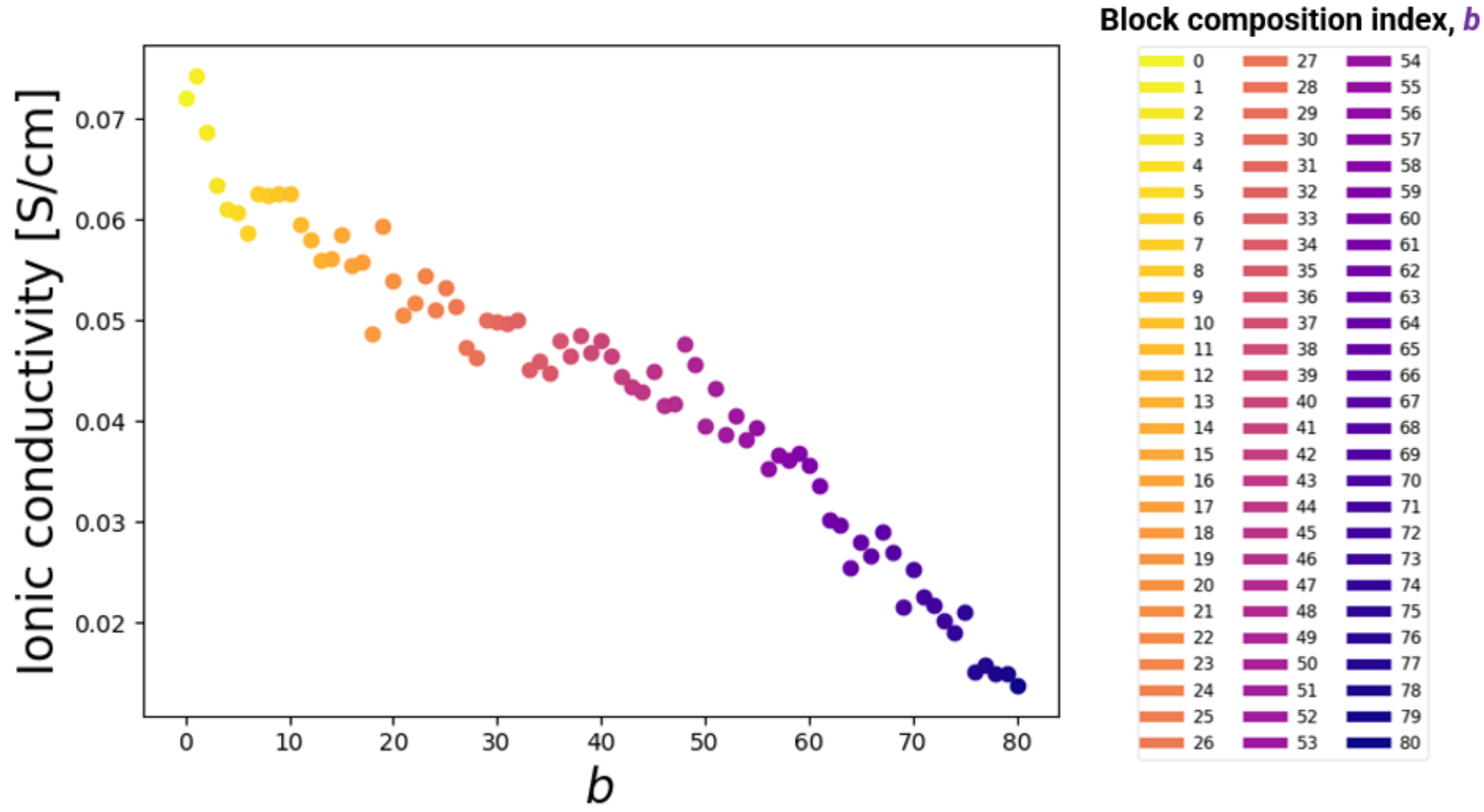

**Figure S5** *Ionic conductivity as a function of block composition.*

**Supporting Note C: Search space reduction**

Here, we describe the size of the search space explored in this study. The size of the search space was reduced by considering the following two points: (i) the dopant configuration in the 16 layers was constructed using three types of four-site blocks, and (ii) the block configuration space was divided into 81 subspaces with different block compositions satisfying the 8YSZ dopant concentration. The search space reduction by considering the block model is summarized in Table S2. The number of possible configurations was reduced to $10^{91}$ from $10^{232}$ by considering the block model. The search space with $10^{91}$ candidates was further divided into 81 subspaces with different block compositions, and the resulting number of possible configurations in each subspace is depicted in Fig. S6.

**Table S2** *Number of sites, number of possible configurations, and bit lengths.*

| Model | All-Atom | Block |
|---|---|---|
| Number of sites, $N_{\mathrm{site}}$ | 800 | 200 |
| Unit | Zr/Y | $Zr_4/Y_2Zr_2/Y_4$ |
| Bit length | 800 | 600 |
| Number of possible configurations (logarithmic scale) | 232 | 91 |

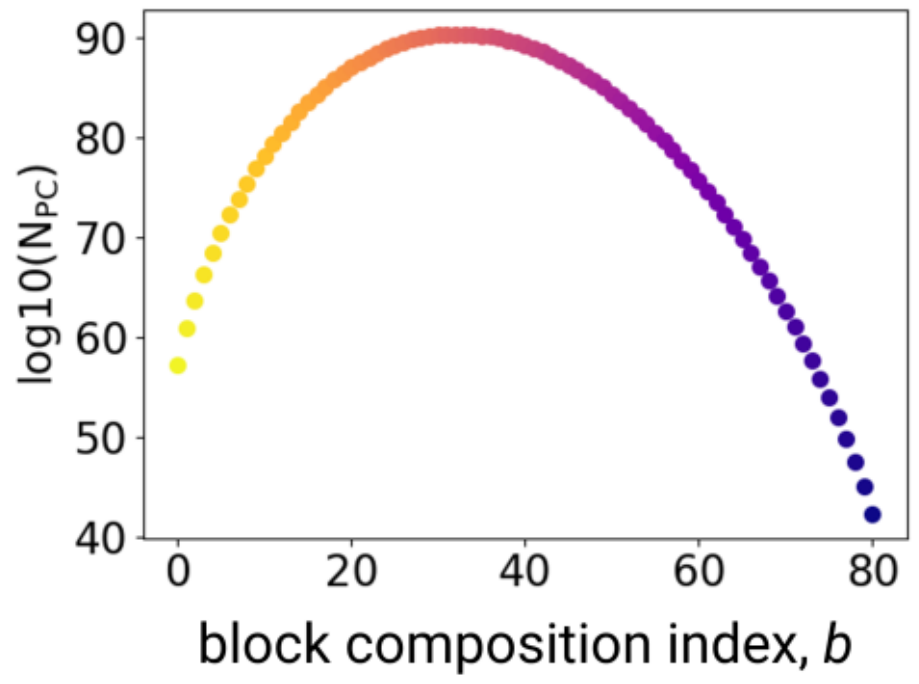

**Figure S6.** *Number of possible configurations,* $N_{\mathrm{PC}}$, *for each block composition.*

**Supporting Note D: Symmetry operations**

Here, we describe the symmetry operations considered for the crystalline system, as illustrated in Fig. 1(c). The system has 10 layers in each of the x and y directions, and 40 layers in the z direction. A weak bias voltage was applied in the z direction, and thus we considered symmetry operations for only the x and y directions. The crystalline system is constructed by repeating the four-site block structure, and thus translational operations of 5 × 5 in both the x and y directions are permitted. Additionally, rotation operation around the $C_2$ axis and mirror operation with respect to the red plane are permitted, as illustrated in Fig. S7(a)-(d). Thus, ultimately, we considered 100 (5 × 5 × 4) symmetry operations in this study.

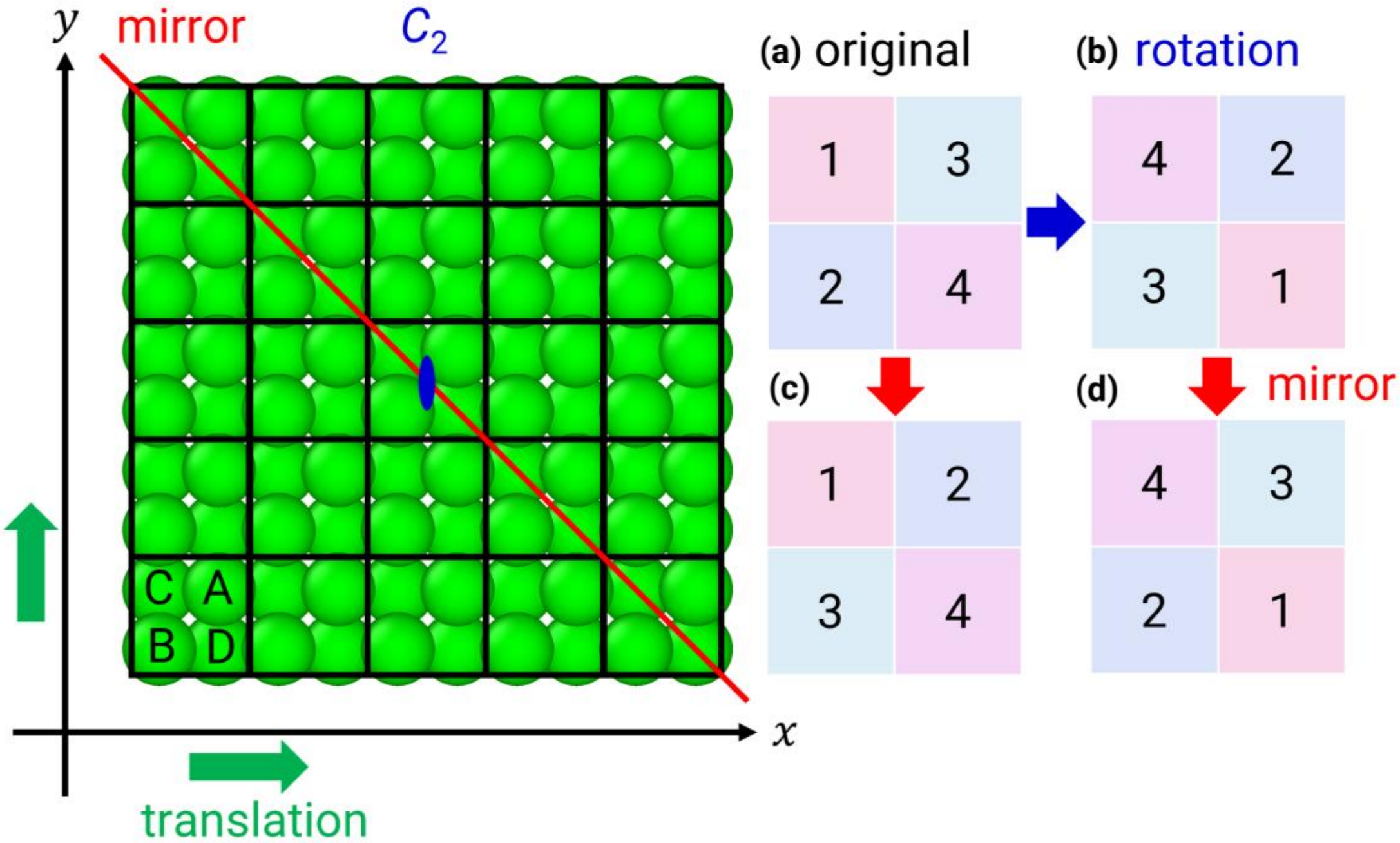

**Figure S7.** *Symmetry operations.* The crystal structure is shown in the x-y plane, with the z direction perpendicular to the plane of the paper. The red line and blue ellipse indicate the mirror plane and the $C_2$ axis, respectively, extending perpendicular to the x-y plane (i.e., along the z direction).